\documentclass[prl,twoside,amsfonts,amssymb,amsmath,titlepage,floatfix,letterpaper,twocolumn,superscriptaddress]{revtex4}
\usepackage{amsmath}
\usepackage{amsfonts}
\usepackage{amssymb}
\usepackage{graphicx}
\usepackage{graphicx,array}
\usepackage{dcolumn}

\newcommand{\ket}[1]{\ensuremath{\vert #1 \rangle}}

\begin{document}

\title{Realisation of Hardy's Thought Experiment}

\author{William T.M. Irvine}
\email[corresponding author: ]{william@physics.ucsb.edu}
\affiliation{Department of Physics, University of
California, Santa Barbara, CA 93106, USA}
\affiliation{Department of Physics, University of Oxford, Parks Road,
Oxford OX1 3PU, United Kingdom}

\author{Juan F. Hodelin}
\affiliation{Department of Physics, University of
California, Santa Barbara, CA 93106, USA}

\author{Christoph Simon}
\affiliation{Laboratoire de Spectrom\'{e}trie Physique, CNRS -
Universit\'{e} de Grenoble 1,  St.\ Martin d'H\`{e}res, France}

\author{Dirk Bouwmeester}
\affiliation{Department of Physics, University of
California, Santa Barbara, CA 93106, USA}


\begin{abstract}
\small We present an experimental realisation of Hardy's thought experiment~[Phys. Rev. Lett. {\bf 68}, 2981 (1992)], using photons. The experiment consists of a pair of Mach-Zehnder interferometers that interact through photon bunching at a beam splitter. A striking contradiction is created between the predictions of quantum mechanics and local hidden variable based theories. 
The contradiction relies on non-maximally entangled position states of two particles.
\end{abstract}

\maketitle

Quantum mechanics poses a challenge to the notion that
objects carry with them values of observables, such as position, that 
both determine the outcomes of measurements and that are local, i.e. uninfluenced 
by events that happen outside the object's backward light cone.
It was first pointed out by 
Bell~\cite{bell} 
that the predicted  correlations between outcomes of measurements on  two spatially
separated systems prepared in an entangled quantum state, were too
strong to be reproduced by any theory based on local `hidden'
variables (LHVs). He formulated an inequality which places a bound on 
the correlations predicted by any such theory, opening the possibility 
of performing experimental tests whose realisation~\cite{bellexpmts} decided in favour 
of quantum mechanics. By considering three particles in an entangled quantum state 
Greenberger, Horne and Zeilinger~\cite{ghz} later proposed a scheme
in which  quantum mechanics and LHV theories predict opposite measurement outcomes, leading to an even stronger contrast between the two. The predictions were verified 
 with polarisation entangled photons~\cite{ghzexp}.

More recently, Hardy formulated a thought experiment~\cite{hardy} that involves only two
spatially separated particles, like in the Bell case, but which leads to a strong contradiction between  the LHV and quantum mechanical predictions  in a  similar  spirit to the GHZ scheme. A crucial feature in Hardy's thought experiment is that the two particles are non-maximally entangled. He later generalised his scheme in a way that could be tested with polarisation entangled photons~\cite{hardygen,dem}.

Here we present an experimental realisation of Hardy's original thought
experiment, using photons. It differs from the polarisation-based scheme in that 
the variables used are paths taken by photons. This makes the contradiction particularly striking, since position is an external variable that translates intuitively to its classical equivalent.

\begin{figure}
\includegraphics[width=\columnwidth]{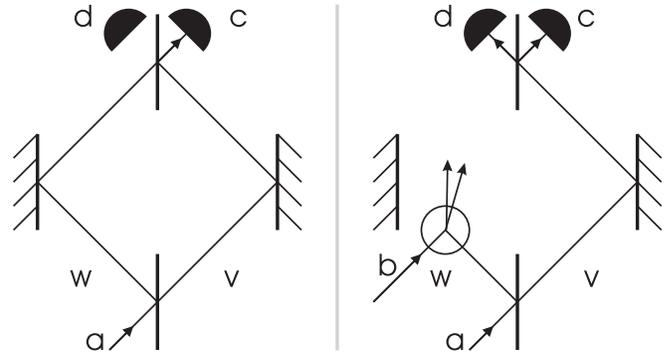}
\caption{Scheme for the implementation of an interaction-free
measurement~\cite{intfree}. {\sc Left:} A Mach-Zehnder interferometer: a particle in mode $a$ 
splits at the first beam splitter and interferes at the second to emerge exclusively  in c. {\sc
Right:} If the interference is disturbed by the presence of a  second particle in  $b$ that collides with the amplitude for the first particle to be in $w$, there will be a probability $p=\frac{1}{4}$ for the first particle to emerge in  $d$. Detection of the particle in $d$ then indicates the presence of a particle in b, without the latter being affected. } \label{fig:intfree}
\end{figure}
The basic building block of Hardy's thought experiment is a Mach-Zehnder interferometer
 for quantum particles (Fig.~\ref{fig:intfree}, Left). The interferometer  is tuned 
so  that particles entering in arm $a$ exit in arm $c$:
\[
\ket{a}\rightarrow \frac{\ket{v}+i \ket{w}}{\sqrt{2}}\rightarrow  \frac{(\ket{d}+i\ket{c})+i (\ket{c}+i\ket{d})}{2}= i \ket{c}.
\]
If the amplitude for the particle in one arm, say $w$, were to be obstructed
by a second particle in arm b that collides with it (Fig.~\ref{fig:intfree}, Right), only the $v$ amplitude would
reach the second beam splitter, and would split into arms $c$ and $d$ with equal amplitude. The detection of a particle in arm $d$ would thus indicate the presence of the obstructing particle without  the latter being affected. For this reason, this effect, first proposed by Elitzur and Vaidman~\cite{intfree}, was named ``interaction free'' measurement. 

\begin{figure}
\includegraphics[width=\columnwidth]{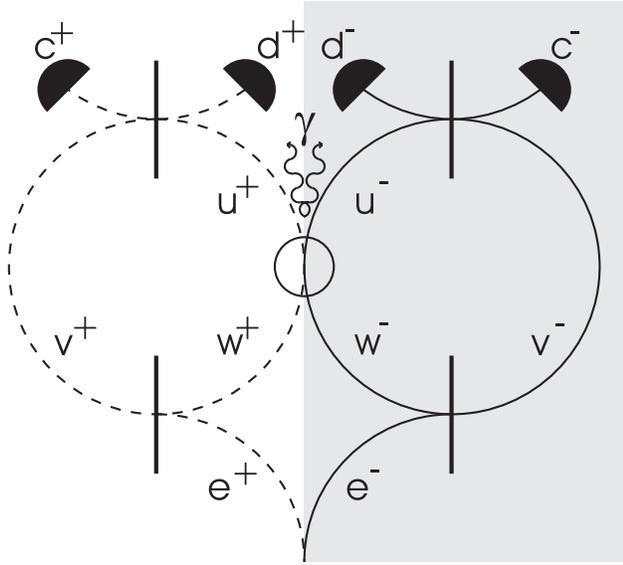}
\caption{Setup for Hardy's original thought-experiment: electrons and positrons enter two interferometers (one on the left, unshaded and one on the right, shaded) simultaneously. If both particles take arms $w$, they annihilate. 
}  \label{fig:hardyth}
\end{figure}
Hardy's original thought experiment (Fig.~\ref{fig:hardyth}) has two interferometers, one for electrons and one for positrons, arranged  in such a way that their $w$ arms intersect. If both the electron and the positron take arms $w$ in their respective interferometers, they will annihilate with certainty to produce gamma radiation: $\ket{w^+}\ket{w^-}\rightarrow \ket{\gamma}$. Therefore the presence of either particle in its $w$-arm will affect the other's interferometer output:
\begin{eqnarray}
&&\ket{e^+}\ket{e^-}  \rightarrow \frac{1}{2} \Big(\ket{v^+}+i\ket{w^+}\Big)\Big(\ket{v^-}+i\ket{w^-}\Big) \nonumber \\
&&\rightarrow \frac{1}{2} \Big[\ket{v^+}\ket{v^-}+ i\ket{u^+}\ket{v^-}+ i\ket{v^+}\ket{u^-}-\ket{\gamma}\Big] \nonumber \\
&&\rightarrow \frac{1}{4}\Big[-3\ket{c^+}\ket{c^-}+i\ket{c^+}\ket{d^-}+i\ket{d^+}\ket{c^-} \nonumber  \\
&&\qquad \qquad \qquad \qquad \qquad \qquad -\ket{d^+}\ket{d^-}-2\ket{\gamma}\Big]. \label{eq:dd}
\end{eqnarray}
The situation can be analysed in terms of two simultaneous interaction-free measurements: From the point of view of the interferometer on the left, a click at $d^+$ implies the presence of the obstructing electron in $u^-$: $[d^+\Rightarrow u^-]$.  Similarly, for the interferometer on the right, a click at $d^-$ implies the presence of the obstructing positron in $u^+$: $[d^-\Rightarrow u^+]$. Indeed every time a click is recorded at $d^\pm$ the other particle is found in $u^\mp$. If we assume the particles are independent (described by LHVs), we conclude that the particles can never emerge simultaneously in  $d^+$ and $d^-$. This would imply that they were in $u^+$ and $u^-$, which cannot occur because of the annihilation process.

A paradox then arises because  sometimes (Eq.~\ref{eq:dd})  the particles {\it do} emerge simultaneously at $d^+$ and $d^-$ ($\mathrm{with\ probability\ } p=\frac{1}{16}$). 
Quantum mechanically, the $\ket{d^+}\ket{d^-}$ term arises in fact from the non-maximally entangled nature~\cite{hardygen}
of the  state just before the final beam splitters $\ket{v^+}\ket{v^-}+ i\ket{u^+}\ket{v^-}+ i\ket{v^+}\ket{u^-}$.


It is instructive to analyse a single run of the experiment from the point of view of different frames of reference. An inertial frame of reference can always be chosen in which one particle, say the positron, reaches a detector before the other reaches the final beam splitter. In that frame upon recording a click at $d^+$ one  can make the   prediction that the electron is in arm $u^-$ with probability $p=1$ since the state of the electron is projected onto \ket{u^-}. Alternatively one can choose a frame moving in the opposite direction and, upon recording a $d^-$ event,  predict with certainty that the positron is in arm $u^+$. Thinking locally, one would then argue that each particle must have travelled in its $w$ arm. 
However by comparing results in the different frames one then runs into a contradiction~\cite{hardy,comments} because had they come from $w^+$ and $w^-$, they would have annihilated. 
 Changing frames in this way allows the paradox to be established for a single setup in which the final measurement is conducted in the $c/d$ basis.

\begin{figure}
\includegraphics[width=\columnwidth]{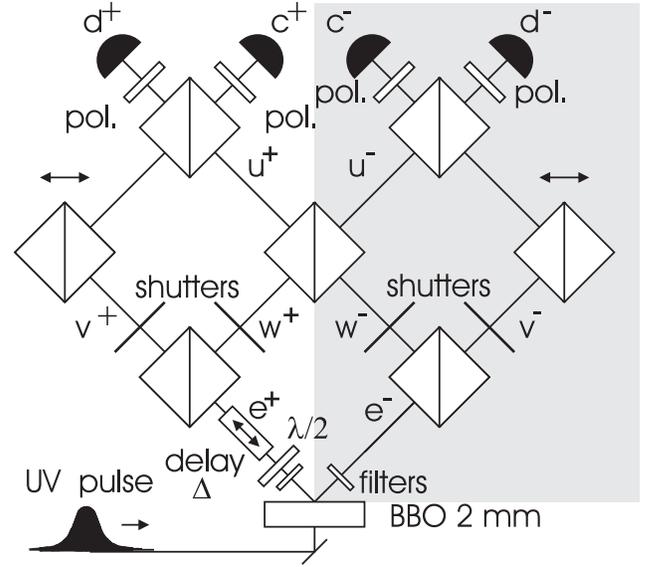}
\caption{Setup for the implementation of Hardy's thought experiment using photons. The two interferometers (left: unshaded, right: shaded) share a central beam splitter, where indistinguishable photons taking arms $w^+$ and $w^-$ bunch. The outermost beam splitters balance the losses from one interferometer to the next through the central beam splitter. Post-selection of the cases in which one photon emerges from each interferometer, leads to a  state identical to that of Hardy's thought experiment (Fig.~\ref{fig:hardyth}).}
\label{fig:apparatus}
\end{figure}
Our scheme to implement the thought experiment, which follows essentially the proposal of Ref.~\cite{hardybunch}, uses indistinguishable photons as a substitute  for the electron and positron and photon bunching at a beam splitter~\cite{hom} as the annihilating interaction. 
The central part of our setup  is a set of  seven beam splitters arranged as in Fig.~\ref{fig:apparatus}. The two interferometers share a central beam splitter where the bunching occurs and can be identified as the sets of four beam splitters on the left (unshaded)  and on the right (shaded). The outermost beam splitters balance the losses through the central beam splitter.  The path-lengths are tuned so that photons entering $e^+$, if not lost through the central or outer beam splitters, emerge exclusively in arm $c^+$. Similarly light entering $e^-$ emerges in $c^-$. In the experiment, pairs of identical photons arrive simultaneously in arms $e^+$ and $e^-$ and enter their respective interferometers: $\ket{e^+}\ket{e^-}\rightarrow
\frac{1}{2}(\ket{v^+}+i\ket{w^+})(\ket{v^-}+i\ket{w^-})$.
As in the electron-positron case, four terms can be identified
corresponding to the four combinations of the paths the two
photons can take.

The $\ket{w^+}\ket{w^-}$ term will bunch at the central beam splitter,  $\ket{w^+}\ket{w^-}\rightarrow\frac{i}{\sqrt{2}}(\ket{2u^+}+\ket{2u^-})$.
This excludes the possibility of detecting a photon leaving each interferometer simultaneously. The absence of such a coincidence click plays an equivalent role to the electron-positron annihilation. The  $\ket{v^+}\ket{v^-}$ term evolves into a superposition of states in which neither, one or both photons are lost through the outermost beam splitters. The cases in which one or both are lost do not give rise to a coincidence click 
and are therefore not counted.  The $\ket{v^+}\ket{v^-}$ term then simply picks up a reduction in amplitude and a change in phase from the reflections: $\ket{v^+}\ket{v^-}\rightarrow-\frac{1}{2}\ket{v^+}\ket{v^-}$.
Finally, the $\ket{v^\pm}\ket{w^\mp}$ terms also evolve into a superposition of states in which one photon is lost through the outermost beam splitter or one photon crosses over to the other's interferometer or the photons end up one in a $v^\pm$ and the other in a $u^\mp$ arm. Post-selection on coincidence counts  gives the evolution: $\ket{v^\pm}\ket{w^\mp}\rightarrow-\frac{1}{2}\ket{v^\pm}\ket{u^\mp}$. Combining these terms, we get the desired post-selected state of the thought experiment:
\begin{equation}
\frac{1}{\sqrt{3}}\Big[\ket{v^+}\ket{v^-} + i\ket{u^+}\ket{v^-} + i\ket{u^-}\ket{v^+}\Big],
\end{equation}
cf. second line of Eq.~(\ref{eq:dd}), dropping the \ket{\gamma} term which does not give rise to a coincidence click.
The paradox is the same: $[d^\pm\Rightarrow u^\mp,\ p(d^+d^-)=\frac{1}{16}>p(u^+u^-)=0]$. 

In practice, neither the bunching nor the implications $[d^{\pm}
\rightarrow u^{\mp}]$ will be perfect. It is therefore necessary to
derive an inequality describing the  predictions of  LHV theories. An LHV theory simultaneously predicts the results for the complementary $c/d$
and $u/v$ measurements for any given value of the hidden
variables. The predicted results on one side must be independent of
the measurement performed on the other side. Probabilities such as
$p(u^+,c^+;u^-,c^-)$ denote the fraction of all hidden variable
values that give the results shown in the brackets for the
respective measurements. These probabilities are not directly measurable, however the values of measurable probabilities, such as $p(u^+;u^-)$, are derived from them according to a simple rule:
\begin{eqnarray}
p(u^+;u^-)=p(u^+,d^+;u^-,d^-)+p(u^+,d^+;u^-,c^-)\nonumber\\
+p(u^+,c^+;u^-,d^-)+p(u^+,c^+;u^-,c^-),
\label{u+u-}
\end{eqnarray}
one adds all possible outcomes for the complementary
measurements on each side. Since probabilities are positive,
this implies:
\begin{equation}
p(u^+;u^-) \geq p(u^+,d^+;u^-,d^-) \label{u+u-ineq}
\end{equation}
 The expression for  $p(d^+;d^-)=p(u^+,d^+;u^-,d^-)+p(u^+,d^+;v^-,d^-)+p(v^+,d^+;u^-,d^-)+p(v^+,d^+;v^-,d^-)$
 allows us to rewrite Eq. (\ref{u+u-ineq}) as:
\begin{eqnarray}
p(d^+;d^-) \leq p(u^+;u^-) +
p(u^+,d^+;v^-,d^-)\nonumber\\
+p(v^+,d^+;u^-,d^-)+p(v^+,d^+;v^-,d^-).
\label{d+d-ineq}
\end{eqnarray}
To bound the last three terms on the right hand side, one can use
the equalities:
\begin{eqnarray}
p(d^+;v^-)=p(u^+,d^+;v^-,c^-)+p(u^+,d^+;v^-,d^-)\nonumber\\+p(v^+,d^+;v^-,c^-)+p(v^+,d^+;v^-,d^-),\nonumber\\
p(v^+;d^-)=p(v^+,c^+;u^-,d^-)+p(v^+,c^+;v^-,d^-)\nonumber\\+p(v^+,d^+;u^-,d^-)+p(v^+,d^+;v^-,d^-),
\label{eqls}
\end{eqnarray}
derived as above, to obtain:
\begin{eqnarray}
p(d^+;v^-)+p(v^+;d^-) \geq p(u^+,d^+;v^-,d^-) \nonumber\\
+p(v^+,d^+;u^-,d^-)+p(v^+,d^+;v^-,d^-).
\end{eqnarray}
Using this inequality in Eq. (\ref{d+d-ineq}) gives the final result:
\begin{equation}
p(d^+;d^-) \leq p(u^+;u^-)+p(d^+;v^-)+p(v^+;d^-), \label{final}
\end{equation}
 which is similar to the
 Clauser-Horne inequality \cite{clauserhorne}.

We now discuss the experimental requirements to violate this inequality.
The quality of the bunching depends on the distinguishability of the photons emerging from the central beam splitter. The parts of their wave-packets which are distinguishable do not bunch and will either both be reflected, both be transmitted or both end up on the same side. The case  in which they are both reflected is equivalent to each particle remaining in its own interferometer and therefore  leads to a $c^+c^-$ click.  The case in which they are both transmitted however, leads to the photons each emerging randomly from the last beam splitters giving an equal amount of $c^+c^-$, $c^+d^-$, $d^+c^-$ and $d^+d^-$ clicks. The implications $[d^\pm\Rightarrow u^\mp]$ remain unaltered since the only way a $d^+$ or a $d^-$ click can arise is from the amplitude in which both photons swap interferometers, which can only have occurred  if the photons emerge in arms $u^+$ and $u^-$. 
Consequently, the quality of the implications depends only on the quality
of the interferometers, which in turn depends
on their alignment and can be made high. By contrast, the
quality of the annihilation  poses a  stronger restriction than might be expected. A
$u^+u^-$ event arising from the  distinguishable amplitudes both swapping interferometers only leads to a $d^+d^-$ click  with a probability $p=\frac{1}{4}$, contributing  four times as many  $u^+u^-$  than $d^+d^-$ events. Even assuming perfectly working interferometers $(p(d^\pm;v^\mp)=0)$, this leads to the requirement that the probability of the photons being distinguishable $p(\mathrm{disting.})$ be less than $\frac{1}{8}=12.5\%$ for the inequality (\ref{final}) to be violated.

In our experimental setup (Fig. 3) a Ti:Sapph mode-locked laser produces
light pulses of 120fs duration, centered at a wavelength of 780nm,
with a repetition rate of 82MHz. The light is passed through a
$\beta$-Barium Borate ($\beta$-BBO) crystal where it is frequency doubled. The frequency
doubled light then arrives at a second, 2mm thick $\beta$-BBO
crystal where it is down-converted~\cite{boyd} to produce pairs of
near-degenerate photons having orthogonal polarization. To make the
photons less distinguishable, two 3nm bandwidth interference filters
were placed in arms $e^+$ and $e^-$, together with a half-wave plate 
at $45^\circ$ in arm $e^+$ to align the polarizations. For a detailed discussion on the effect of filters on bunching in this type of system, see~\cite{walmsley}.
The light then passes through the setup and is detected at $c^+$,$c^-$,$d^+$ and $d^-$. The outermost beam splitters are mounted on piezo-electrically driven translation stages used to tune the length of the $v$-arms of the interferometers.

We measure simultaneous clicks between detectors on the left  and on  the right ($c^+c^-$,$c^+d^-$,$c^-d^+$,$d^+d^-$).
Coincidence logic allows the distinction of genuine two-photon events from random dark counts. To measure in the $c/d$ or $u/v$ bases we close the shutters in arms $v$ and $w$ as appropriate. For example, if the $w^-$ arm is blocked, then all the photons reaching $c^-$ and $d^-$ must have come from arm $v^-$. Thus to measure, say, the
number of $d^+v^-$:  $N(d^+u^-)$ we measure $N(d^+c^-)+N(d^+d^-)$
with the shutter in arm $w^-$ closed. 

For a fair measurement of the contradiction $N(d^+d^-)>N(u^+u^-)+N(d^+v^-)+N(v^+d^-)$ we need to make sure that the right hand side is not under-estimated by our measurement technique. To do this we measured the rates $N(d^+d^-)$, $N(c^+d^-)$, $N(d^+c^-)$ and $N(c^+c^-)$ under all combinations of blocking arms $v^\pm$ and $w^\pm$ and ensured the efficiencies were all within 10\% of each other, with the $d^+d^-$ efficiency  less than the others in all configurations. 
\begin{figure}
\includegraphics[width=\columnwidth]{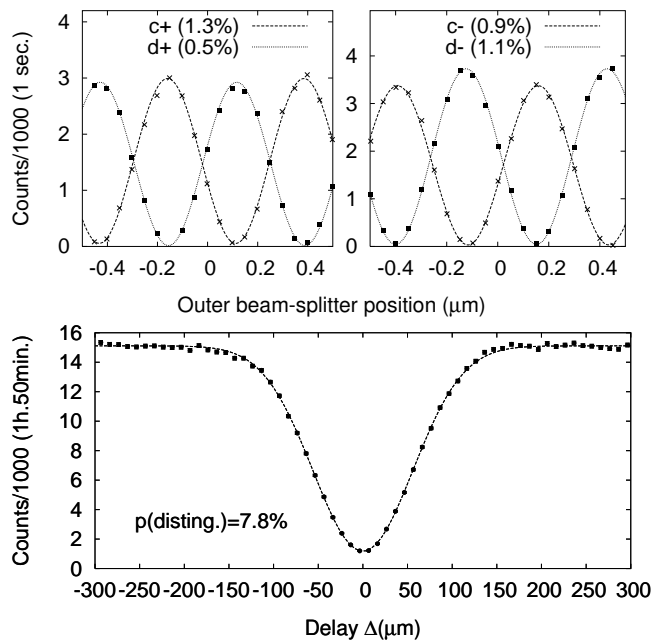}
\caption{{\sc Top:} Interference fringes that
characterize the output of each  interferometer at the time of the
experiment. The probabilities that a photon emerges in the  $d$-arm deduced from the fits are
given in the plots. {\sc Bottom:} $N(u^+u^-)$ as a function of
delay $\Delta$. The probability that the photons are distinguishable 
$p(\mathrm{disting.})$ at zero delay is given by
$p(\mathrm{disting.})=N(u^+u^-)_{\Delta=0}/N(u^+u^-)_{\Delta>>0}=8\%$.}
\label{fig:quality}
\end{figure}
Fig.~\ref{fig:quality} shows the quality of the interferometers and of the bunching  at the time of the
experiment. The probability of getting a
$d^+d^-$ click from two photons accidentally emerging in the $d$ arms
is less than 0.55\%. The quality of the bunching is above the threshold required to measure a violation ($p(\mathrm{disting.})=8\%<12.5\%$).

\begin{figure}
\includegraphics[width=\columnwidth]{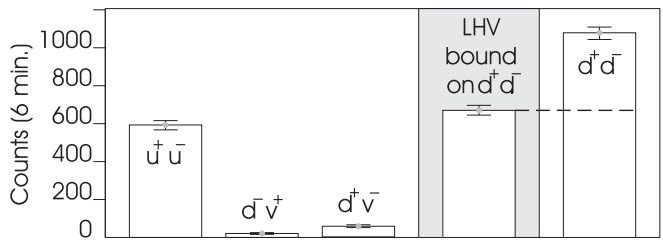}
\caption{The three measurements on the left $N(u^+u^-)$, $N(d^+v^-)$ and $N(v^+d^-)$ lead to an LHV bound on $N(d^+d^-)$ (Eq.~\ref{final}), shown in the shaded area. The measured $N(d^+d^-)$ violates this by 12 standard deviations.}
\label{fig:results}
\end{figure}
Fig.~\ref{fig:results}  shows a comparison between the measured $N(d^+d^-)$ and the LHV bound on it. The bound is given by the sum of the measured $N(d^+v^-)$, $N(d^-v^+)$ and $N(u^+u^-)$.
We find a violation of the LHV inequality (Eq.  \ref{final}) by 12 standard deviations. The violation is consistent with the quantum mechanical predictions based on the probability of  bunching and the detection efficiencies in our setup. 


In conclusion, we have implemented Hardy's thought experiment consisting of two interacting Mach-Zehnder interferometers, demonstrating the contradiction between quantum mechanics and LHV theories in a striking way. It should be noted that the concept of creating entanglement by influencing a single photon interferometer with another photon also plays a crucial role in optical approaches to quantum logic gates~\cite{cnot}.


We would like to acknowledge Ian Walmsley for useful discussions, support from NSF 
grants 0304678 and 0404440, and Perkin-Elmer regarding the SPCM-AQR-13-FC detectors. W.I. acknowledges Elsag s.p.a. for support under MIUR's grant n.67679/L.488/92. J.H. acknowledges support from Lucent Technologies CRFP.

\end{document}